\documentclass[%
 reprint,
 amsmath,amssymb,
 aps,
]{revtex4-2}

\usepackage{graphicx}
\usepackage{dcolumn}
\usepackage{bm}

\begin{document}

\preprint{APS/123-QED}

\title{Exactness of Symmetry-Broken Self-Interaction Correction in the Strongly-Correlated or Classical Limit: Harmonium as a Demonstration\\}

\author{Cody H. Woods$^{1}$}
\author{Yunzhi Li$^{2}$}
\author{Wenqing Yao$^{3}$}
\author{Chen Li$^{3}$}
\author{John P. Perdew$^{1}$}

\affiliation{$^{1}$ Department of Physics and Engineering Physics,
Tulane University, New Orleans, Louisiana 70118, USA}

\affiliation{$^{2}$Key Laboratory of Theoretical and Computational Photochemistry, Ministry of Education, College of Chemistry, Beijing Normal University, Beijing 100875, China}

\affiliation{$^{3}$ Beijing National Laboratory for Molecular Sciences,
College of Chemistry and Molecular Engineering,
Peking University, Beijing 100871, China}

\date{\today}

\begin{abstract}
Strong electron correlation is an important challenge to both wavefunction and density functional theory. It has been argued that the Perdew-Zunger 1981 self-interaction correction to any density functional approximation, after symmetry breaking, can correctly describe the ground-state energy in the strongly-correlated limit in which each electron is described by a highly localized and non-overlapping one-electron spin orbital. It has also been argued that the classical limit, in which Planck’s constant tends to zero, is the strongly-correlated limit of quantum mechanics, where standard density functionals fail badly, as demonstrated by the exactly-solvable problem of harmonium (two Coulomb-interacting electrons bound by a spherically-symmetric harmonic-oscillator external potential). Here we combine these two ideas and demonstrate that, for harmonium, symmetry-broken self-interaction correction is exact in the limit where Planck’s constant tends to zero, and usefully accurate for all values between 0 and the physical value (1 in atomic units). Our results also strongly suggest that the Planck-constant-dependent symmetric ground-state density can be restored by spherical averaging of the broken-symmetry density.
\end{abstract}

\maketitle


Density functional theory (DFT) remains the most widely-used method for electronic structure calculation because of its balance between accuracy and computational cost. DFT typically employs a single Slater determinant of self-consistent spin orbitals, and finds the exchange-correlation energy from a functional of the density. Through the Hohenberg–Kohn theorems \citep{hohenberg_inhomogeneous_1964} and the Kohn–Sham formalism \citep{Kohn1965}, DFT is formally exact for the ground-state energy and density of real electrons in a static external potential. In practice, however, one must rely on exchange–correlation (XC) functional approximations for computational feasibility. Common density functional approximations or DFAs (including the local spin density approximation (LDA) \citep{perdew_self-interaction_1981}, the generalized gradient approximation of Perdew, Burke, and Ernzerhof (PBE) \citep{Perdew1996}, and the meta-generalized gradient approximation r2SCAN \citep{Furness2020,Sun2015}) have achieved remarkable success across a range of systems. These approximations, however, still struggle in many cases and can exhibit large errors \footnote{For a discussion on some common DFT challenges, see \citep{Cohen2008, Cohen2012}}.

An important source of DFA error is the self-interaction error (SIE), which arises because approximate exchange-correlation functionals do not completely cancel the Hartree-based Coulomb interaction of an electron with itself. Manifestation of SIE in a many-electron situation include excessive delocalization, inaccurate dissociation behavior, and failures in strongly-correlated systems \citep{Cohen2012, Cohen2008}. The Perdew-Zunger self-interaction correction (PZ-SIC) \cite{perdew_self-interaction_1981} was introduced to make a functional exact for all one-electron densities.

Strong electron correlation can be problematic for both wavefunction theory and ground state DFT. Standard approximate functionals are often limited to normal correlation, at least in the absence of symmetry breaking. In wavefunction theory, strong correlation calls for a heavy weight on more than one (and sometimes very many) Slater determinants. Li and Li \citep{li_exact_2025} have argued that, for any system with a Hamiltonian that has a classical analog and is bounded from below (like real electrons in an external pseudopotential), the classical limit (where $\hbar\rightarrow 0$ or $m \rightarrow \infty$) is also the strongly-correlated limit. (Note that spin-spin interaction, which has no classical analog, is not discussed here.) They have further illustrated this for the model problem of harmonium \citep{li_exact_2025,yao_formally_2024}, for which they found the exact ground-state energy and density at any value of $\hbar$. Finally, they have assigned an effective value of $\hbar$ (in atomic units where the true value is 1) to each ground-state molecule in a set, based on the ratio of the natural-orbital occupation numbers just above and just below the Fermi level. The lowest effective $\hbar$ they found, around 0.2, was for the strongly-correlated dimers $C_2$ and $Cr_2$. The strong-correlation limit can also be reached by holding the electron density fixed and letting the square of the electron charge approach infinity \citep{Seidl2007}. 

In ground-state DFT, there are many examples, including  $C_2$, where a calculation with a sufficiently-reliable DFA captures the energy of strong correlation by spontaneously breaking the symmetry of the electron density or spin density \citep{Perdew2023SymmetryBreaking,PRS,Z,P2}. The structure of Kohn-Sham theory with an approximate XC functional is such that, for a system of electrons localized in non-overlapped spin orbitals, its only error is the sum of the SIE for each orbital density. Perdew \citep{Perdew2025SCAN} has argued that, because PZ-SIC (applied to any approximate density functional) is exact for any collection of non-overlapped and spin-polarized one-electron orbitals, it is also exact in the broken-symmetry strongly-correlated or classical limit in which electrons come to rest at a configuration that minimizes the total potential energy. (This exactness holds even when the external potential has no symmetry.) We will demonstrate this here for two-electron harmonium, where broken-symmetry PZ-SIC works well for all values of $0 < \hbar < 1$. In ground states of more than two electrons, PZ-SIC needs to be scaled-down locally to recover both the one-electron and the uniform-density limits, and this has so far been achieved reliably only when the functional being corrected is the local spin-density approximation \citep{Shahi2026Local}), in which the Hartree and XC energy densities are in the same gauge. A formal argument \citep{P2} suggests that only a functional that is reliable for normally-correlated states at a given density (including one-electron states) can be reliable for strongly-correlated ones via symmetry breaking. While approximate functionals without SIC can often lower and improve the total energy by spontaneous symmetry breaking, we will see that the harmonium ground state is a case where this can only happen after symmetry-preserving SIC. Symmetry-breaking is a powerful concept even in small systems, but it is almost unavoidable in extended ones where symmetry breaking is widely observed in experiments. A beyond-DFT treatment of a solid is often a quantum cluster theory (such as the dynamical mean field theory), in which a small cluster of atoms is explicitly correlated while its environment is treated in mean-field-like Kohn-Sham DFT. Section II.F of Ref. \citep{Maier2005} discusses how to take account of symmetry breaking in a quantum cluster theory.

In the harmonium model (often, Hooke’s atom \citep{taut_two_1993}), two electrons repel each other via the Coulomb interaction in the Hamiltonian below:
\begin{eqnarray}
\hat{H}= \sum_{i=1}^2 (-\frac{\hbar^2}{2} \nabla_i^2+\frac{1}{2} \omega^2r_i^2) +\frac{1}{|\bf{r_1}-\bf{r_2}|},
\end{eqnarray}
where $\omega$(=0.373 atomic units here as in Ref. \citep{li_exact_2025}) establishes the confining potential strength and our tunable Planck's constant $\hbar$ controls the balance between the quantum delocalization and classical localization. In the large $\hbar$ limit (or at high $\omega$), the system approaches weak correlation, where kinetic energy dominates and electrons become delocalized.

In the $\hbar \rightarrow 0$ limit, harmonium approaches a classical configuration where the electrons localize at finite separation. This localization is energetically favorable because it minimizes the sum of the Coulomb and the harmonic confinement potential energies. Here, the electrons in the system behave like point charges in the harmonic trap. The energy associated with the ground state is the sum of all potential energies,

\begin{eqnarray}
    U(r) = \omega^2 r^2 +\frac{1}{2r}.
\end{eqnarray}

Minimizing this energy with respect to r yields an optimal radius of 

\begin{eqnarray}
    r_0 =(2\omega)^{-2/3}=1.216.
\end{eqnarray}
Plugging into $U(r)$, we obtain the following classical two electron energy:

\begin{eqnarray}
    U_{\mathrm{cl}}=3*2^{-4/3}\omega^{2/3} \approx 0.6169 \ \text{Ha (for} \ \omega = 0.373).
\end{eqnarray}

In contrast to the classical broken-symmetry picture, the quantum description restores the spherical symmetry of the Hamiltonian. There is high degeneracy and thus strong correlation in the spherically-symmetric ground-state wavefunction when $\hbar \rightarrow 0$. As $\hbar$ increases from 0, quantum zero-point energies of the classical normal modes of oscillation grow, creating additional energy proportional to $\hbar\omega$.

\begin{figure}[htbp]
  \centering
  \includegraphics[width=\linewidth]{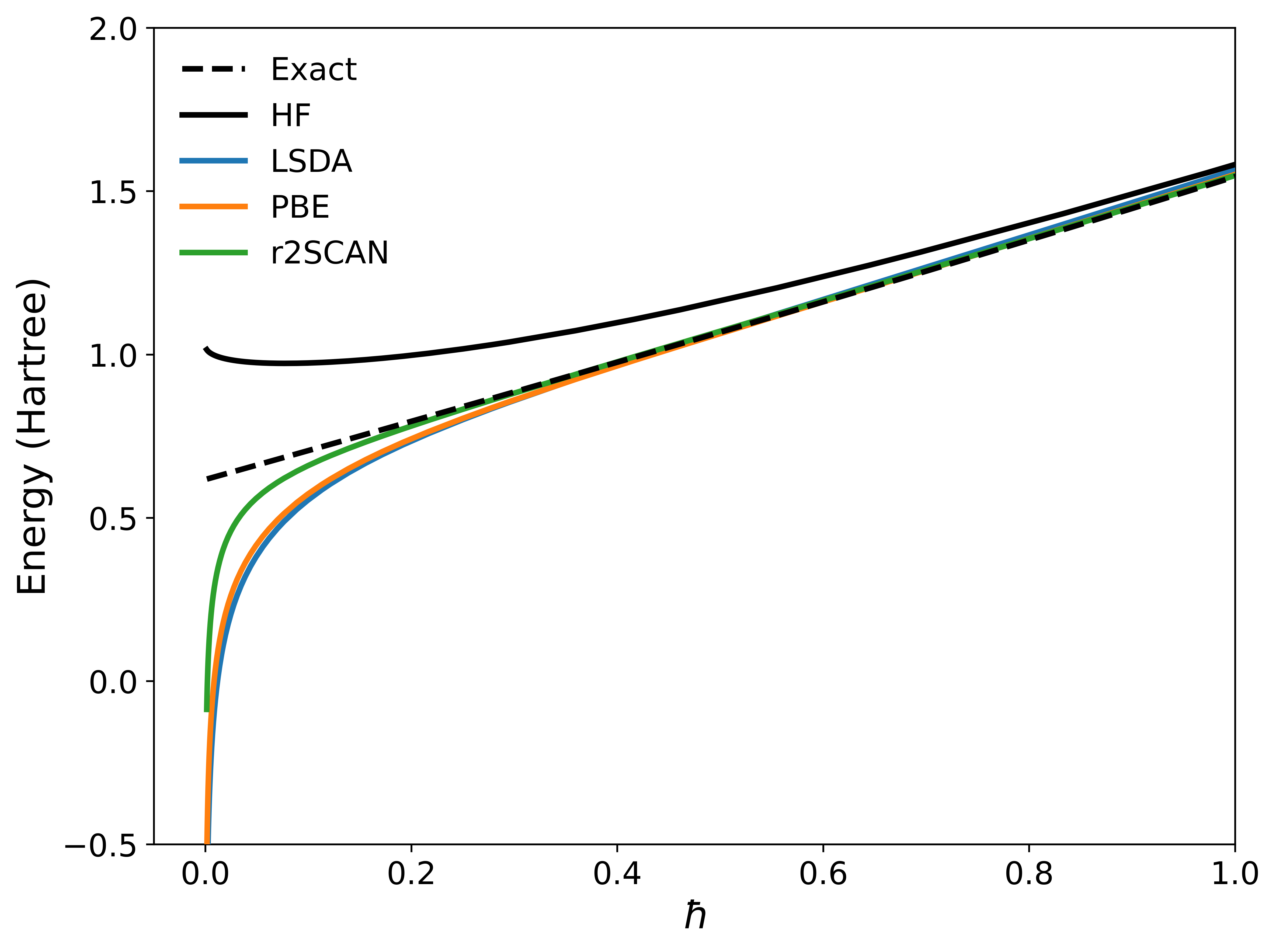}
  \caption{$E$ vs $\hbar$ for harmonium model without SIC, evaluated on the exact density. The black dashed line is the exact energy.}
  \label{fig:NoSIC_E_vs_hbar}
\end{figure}

Figure \ref{fig:NoSIC_E_vs_hbar} is the analog of Fig 8. of Ref. \citep{li_exact_2025}, constructed in the same way: The approximate functionals are evaluated on the exact or nearly exact density. (Self-consistent calculations present technical difficulties that we plan to address in future work, but are not needed to establish our main conclusions.) To make Fig. \ref{fig:NoSIC_E_vs_hbar}, we used the exact ground-state densities (Eq. S6) for $\hbar>0.01$, and the simpler but nearly-exact semi-classical harmonic oscillator densities (Section S1.2 of the SI) for $\hbar\leq0.01$. Because the XC energy functional (more precisely its correlation part) itself must depend on $\hbar$, an exact density scaling (presented in our SI) is used to take account of this usually-implicit dependence of the DFA on $\hbar$. The Hartree-Fock approximation includes exact exchange and no correlation. At  $\hbar=0$, it misses the strong correlation energy (-0.38 Ha), and at $\hbar=1$ it misses the much smaller normal correlation energy. For $\hbar$ between 0.4 and 1, Fig. \ref{fig:NoSIC_E_vs_hbar} shows good agreement of the standard DFAs with the exact energy. LSDA and PBE make larger energy errors than r2SCAN, but all three become improperly negative or even diverge to $-\infty$ as $\hbar\rightarrow0$. For LSDA, the divergence is proportional to $\hbar^{-1/6}$ \citep{li_exact_2025}. All three DFAs are single integrals over three-dimensional space. In LSDA, the integrand depends only on the local density. In the PBE generalized gradient approximation (GGA), the gradient of the density is also employed. r2SCAN further includes the kinetic energy density of the occupied orbitals.
Because the DFA energy errors are all negative, they cannot be improved by spontaneous symmetry breaking that would lower the energy. This is different from the situation encountered in many real strongly-correlated systems, including $C_2$, where SCAN works through spontaneous symmetry-breaking \citep{Perdew2023SymmetryBreaking}. This difference arises from the fact that the harmonium density in the classical limit concentrates on the surface of a sphere of radius $r_0$, a highly noncompact arrangement for which the DFA exchange energies are far too negative, while the valence density of $C_2$ is compact. These results are consistent with the formal theory of DFA symmetry-breaking in Ref. \citep{P2}: A necessary condition for a DFA to capture the energy of strong correlation is that the DFA must be reliable for a normally-correlated state, as LSDA, PBE, and r2SCAN without SIC cannot be for a highly noncompact density.

Thus, standard DFAs fail in the classical limit as $\hbar\rightarrow0$ \citep{yao_formally_2024,li_exact_2025}. The dominant source of DFA error for two-electron harmonium is self-interaction error (SIE). In principle, each electron should interact only with the other, but approximate exchange–correlation functionals imperfectly cancel the self-Hartree energy of each electron. This error is especially severe in the semiclassical regime, where the density is strongly noncompact.
 
The Perdew-Zunger self-interaction correction (PZ-SIC) \citep{perdew_self-interaction_1981} removes one-electron self-interaction on an orbital-by-orbital basis.

\begin{align}
E_{\mathrm{xc}}^{\mathrm{PZ\text{-}SIC}}[n_\uparrow,n_\downarrow]
&=
E_{\mathrm{xc}}^{\mathrm{DFA}}[n_\uparrow,n_\downarrow]
\nonumber\\
&\quad
-
\sum_{i\sigma}
\left[
U[n_{i\sigma}]
+
E_{\mathrm{xc}}^{\mathrm{DFA}}[n_{i\sigma},0]
\right].
\end{align}
Here, the $n_{i\sigma}$ are occupied localized spin densities, typically found by unitary transformation of the occupied Kohn-Sham spin orbitals $\phi_{i\sigma}$, $U[n_{i\sigma}]$ is the standard double-integral Hartree self-interaction energy, and $E_{xc}^{\mathrm{DFA}}[n_{i\sigma},0]$ is the exchange-correlation energy of the DFA evaluated for a one-electron spin density. The symmetry-preserving spin orbital densities for harmonium are uniquely-defined: $n_{i\sigma}(r)=n_{\sigma}(r)=\frac{n(r)}{2}$. The spin orbitals are orthogonal in spin space, not in position space. Since we have two electrons in the ground state, the meta-GGA parameter $\alpha_{\sigma}$ is zero everywhere, so there is no need to scale down the SIC locally to preserve the correct uniform-density limit of the DFAs, as in \citep{Shahi2026Local}. For one- and two-electron ground states in spin-independent external potentials, PZ-SIC is exact for the exchange energy, so any remaining errors are from correlation \citep{perdew_self-interaction_1981}. 

 \begin{figure}[t]
  \centering
  \includegraphics[width=\linewidth]{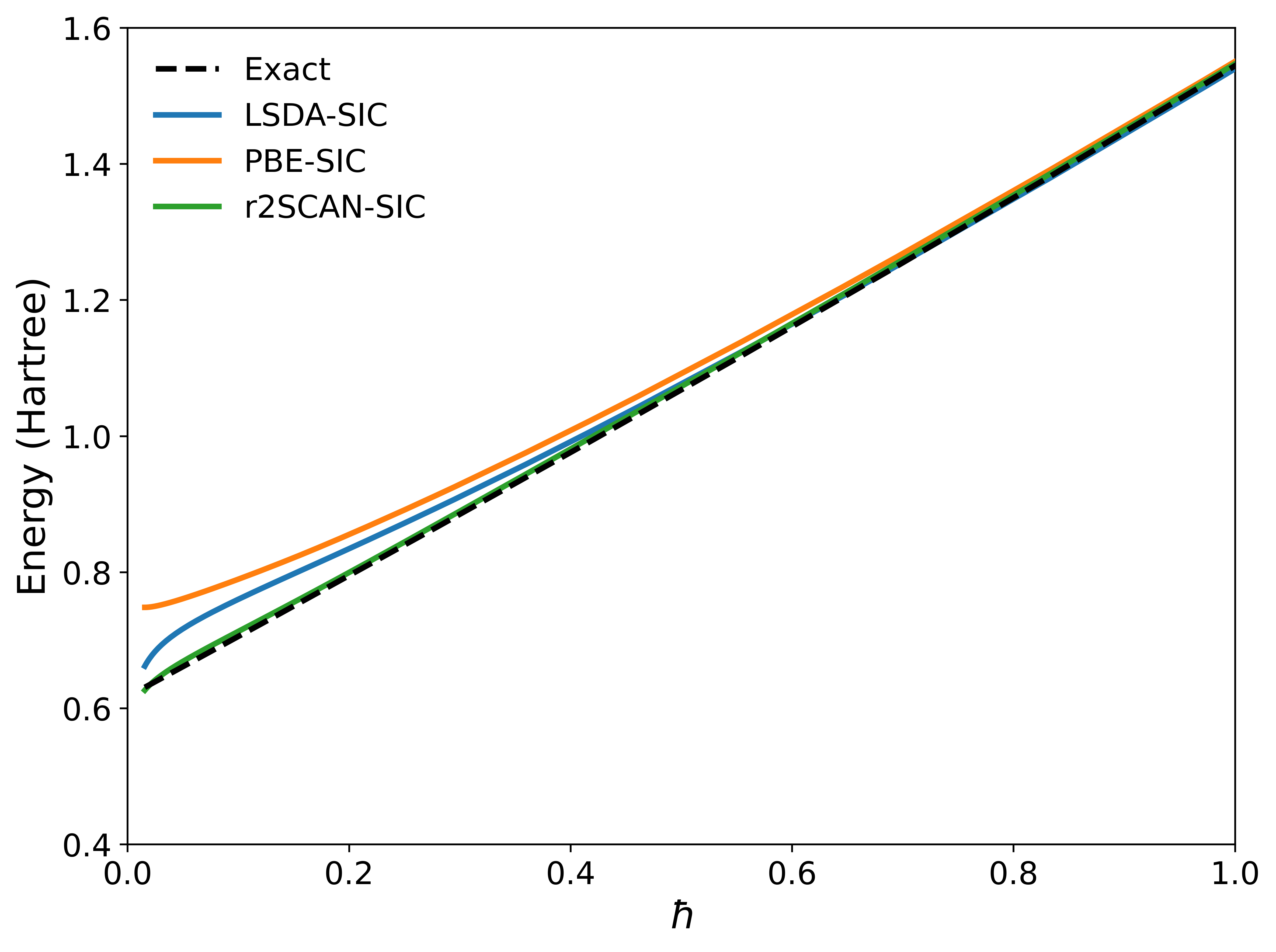}
  \caption{Total energy vs.\ $\hbar$ for LSDA, PBE, r2SCAN SIC variants (symmetry-unbroken), evaluated on the exact density.
  The plot covers the range $0.015< \hbar < 1$. Hartree-Fock, which is already self-interaction-free, remains as in Fig. \ref{fig:NoSIC_E_vs_hbar}.}
  \label{fig:energy_hbar_zoom}
\end{figure}

Figure \ref{fig:energy_hbar_zoom}  highlights the performance of the symmetry-unbroken PZ-SIC approach relative to the exact energy. The energy is evaluated with exact or nearly exact symmetric ground state densities $n(r)$. In the plotted range, the SIC-corrected energies (LSDA-SIC, PBE-SIC, r2SCAN-SIC) track the exact energy more closely than their uncorrected DFA counterparts. r2SCAN is remarkably accurate.  

Furthermore, for all of the DFA functionals,  subtraction of self-interaction produces energies that are non-divergent and above the exact energy. Applying SIC in a symmetry-preserving (restricted) fashion substantially improves the description of harmonium in the strongly-correlated limit. Residual deviations from the exact energy from LSDA-SIC and PBE-SIC, however, remain at low $\hbar$. Importantly, the symmetry-unbroken DFA-SIC errors in the plotted range are all positive, opening the door to spontaneous symmetry breaking.

To investigate this possibility, we construct a variational symmetry-broken ansatz consisting of two localized gaussian orbitals centered at $\pm R$ along a common axis:

\begin{eqnarray}
\phi_L(\mathbf r) = \left(\frac{\alpha}{\pi}\right)^{3/4}
e^{-\alpha |\mathbf r+\mathbf R|^2/2},\\
\phi_R(\mathbf r) = \left(\frac{\alpha}{\pi}\right)^{3/4}
e^{-\alpha |\mathbf r-\mathbf R|^2/2},
\end{eqnarray} 
with corresponding spin densities
\begin{eqnarray}
[
n_\uparrow(\mathbf r)=|\phi_L(\mathbf r)|^2,
\qquad
n_\downarrow(\mathbf r)=|\phi_R(\mathbf r)|^2
].
\end{eqnarray}

These gaussian functions are an arbitrary but natural choice because they are the exact ground state solutions of a harmonic potential. In the strongly-correlated limit, each electron localizes near a classical equilibrium position. In this regime, the total potential is locally quadratic, so the ground state orbital is expected to be approximately gaussian. In fact, we have modulated the trial spin densities of Eq. 8 to have the radial dependence of the exact symmetric density and an angular dependence controlled by the single variational parameter $B=2\alpha R$, as explained in the SI.

\begin{figure}[t]
  \centering
  \includegraphics[width=\linewidth]{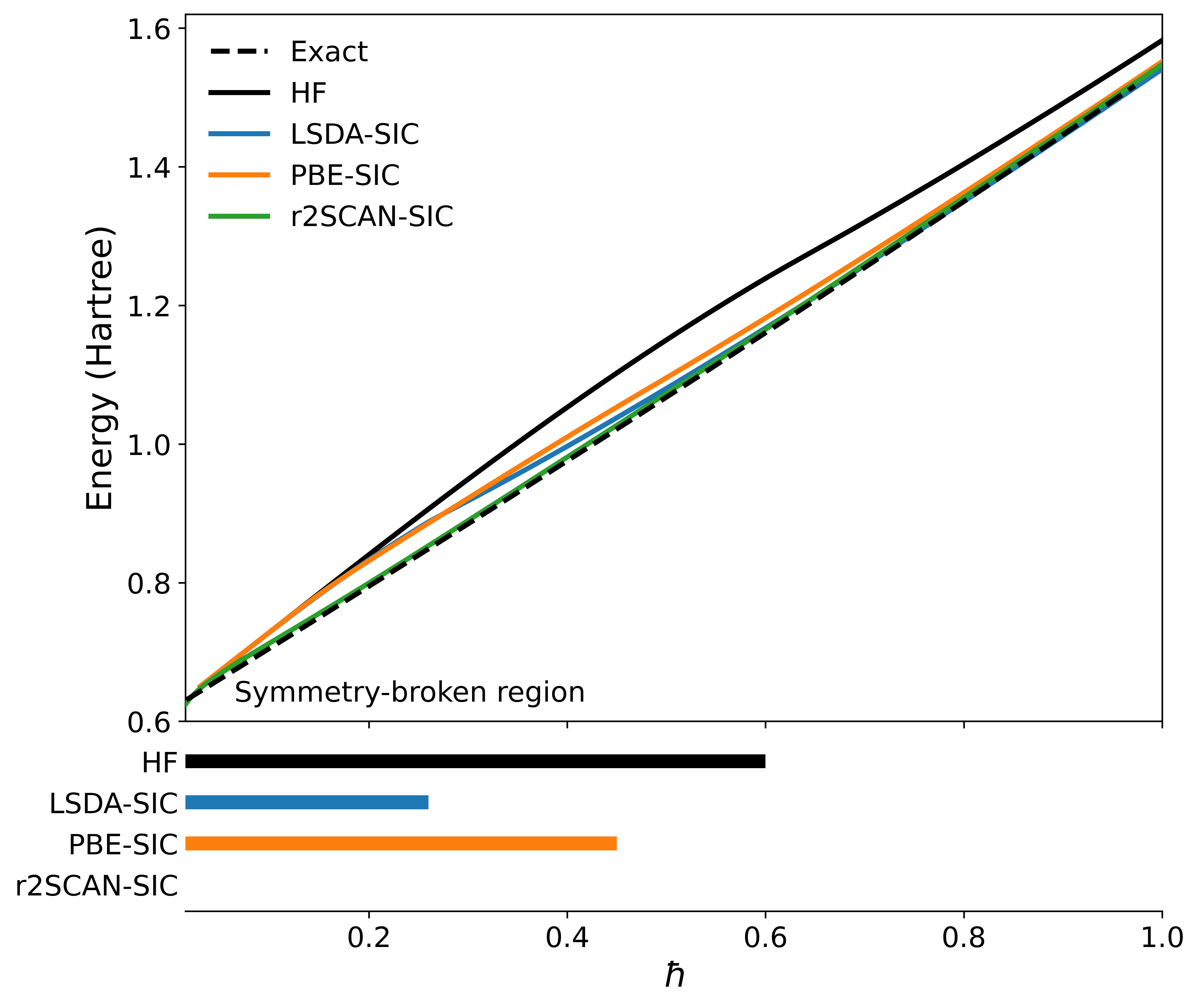}
  \caption{Implementation of SIC (with spontaneous symmetry-breaking), evaluated on the modulated trial spin densities of Eq. 8. For a given functional and $\hbar$, the plotted energy is the lower of the symmetry-broken and symmetry-unbroken energies. The black dashed line is the exact energy. Symmetry-broken HF, LSDA-SIC, PBE-SIC, and r2SCAN-SIC yield the same exact energy in the strongly-correlated limit $\hbar\rightarrow0$. The exactness of symmetry-broken HF for the energy in the strongly-correlated or stretched-bond limit of $H_2$ has long been known. In the normally-correlated limit $\hbar\rightarrow1$, the HF energy misses the normal correlation that the DFAs capture. Color bars at the bottom show the range over which the symmetry-broken solution has the lower energy.}
  \label{fig:variable_alpha_SIC}
\end{figure}

Figure \ref{fig:variable_alpha_SIC}
shows the variationally-optimized and spontaneously symmetry-broken energies as a function of $\hbar$. As $\hbar\rightarrow0$, the noninteracting kinetic energy becomes negligible, and the broken-symmetry occupied orbitals become strongly localized and nonoverlapping. Any spurious interaction of such an orbital with itself is removed by SIC. The remaining energy consists only of the external potential energy and the Hartree interaction between different localized electrons, which reduces exactly to the classical electrostatic energy. The optimized localization parameter $B=2\alpha R$ grows as $\hbar\rightarrow0$ (see figures in the SI). No symmetry-breaking is found for $\hbar\geq0.65$. 

\begin{figure}
    \centering
    \includegraphics[width=1\linewidth]{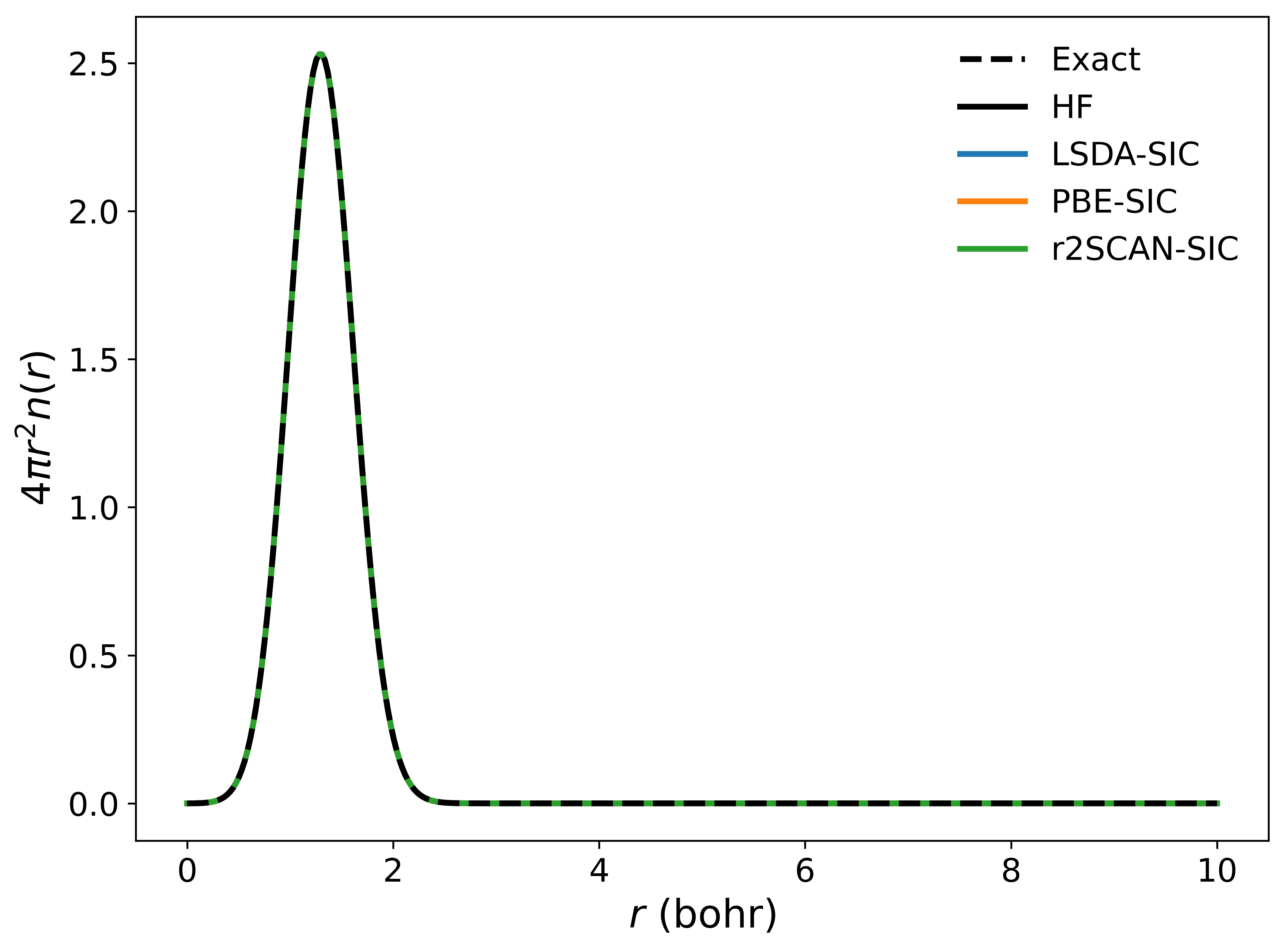}
    \caption{Radial probability density for $\hbar=0.1$. The physically-motivated broken-symmetry spin densities were constructed to have the same spherical average as the exact symmetric density.}
    \label{fig:Radial_Prob_Dens}
\end{figure}

The optimized broken-symmetry densities obtained from the variational PZ-SIC calculations were spherically-averaged over all orientations. Figures \ref{fig:Radial_Prob_Dens} and S3 compare these averaged densities with the exact density at $\hbar=0.1$. Although the underlying localized densities differ slightly in their optimal parameters and total energies, their spherical averages reproduce by construction both the radial structure and overall shape of the exact density. They also eradicate spin polarization and restore the singlet spin symmetry. The symmetry-broken density in harmonium corresponds to a localized configuration that would be observed in a sufficiently short-time measurement, whereas the exact density would correspond to an infinite-time average over all of the equivalent orientations. In this sense, the exact density is not inconsistent with localization; rather, it represents the symmetry-restored average of a family of localized one-electron densities. This result is clear in the classical limit: The simplest quantum description of the classical limit of the harmonium ground state is a Slater determinant of two point-like broken symmetry spin orbitals on an axis through and equidistant from the center. But a general ground state wavefunction is a superposition of such degenerate non-overlapped states, and the symmetric ground state is an equally-weighted superposition of them all.

In summary, as stated in Ref. \citep{li_exact_2025}, the strong-correlation limit of the quantum ground state for a system of particles (whose Hamiltonian has a classical analog and is bounded from below) is the classical limit in which Planck’s constant and the kinetic energy go to zero and the particles localize at positions of minimum total potential energy. Without SIC, the DFAs LDA, PBE, and r2SCAN all fail for the strongly-correlated limit of harmonium. The highly noncompact density in this limit makes the exchange energy and the total energy too low, so these errors cannot be corrected by spontaneous symmetry breaking. Symmetry-preserving SIC yields exact exchange and so raises these energies, opening the door to the benefits of spontaneous symmetry breaking (needed by LSDA-SIC and PBE-SIC in harmonium). Over the range $0.015<\hbar<1$ of any possible relevance to real materials, symmetry-unbroken r2SCAN-SIC, evaluated on the exact density, describes harmonium’s energy with high accuracy, and symmetry-broken r2SCAN-SIC on the variational trial spin densities is comparably accurate. The symmetry breaking in Fig. 3 does not produce visible slope changes in the plotted curves.

As predicted in Ref. \citep{Perdew2025SCAN}, symmetry-broken SIC applied to any DFA yields the exact energy in the strongly-correlated limit.We have demonstrated this numerically for harmonium, where symmetry-broken PZ-SIC is exact in the strong-correlation limit and remains accurate over the range $0<\hbar<0.1$. Since strong correlation also appears in ground states with more than two electrons, this result motivates the search for a reliable scaled-down SIC to approximate density functionals beyond the local spin density approximation, extending the work of \citep{Shahi2026Local}. In harmonium, spherical averaging of the localized symmetry-broken densities accurately reproduces the symmetry-preserving exact density. This result supports the time-centered interpretation of symmetry breaking \citep{A,PRS,P2,Z}), in which the localized density corresponds to a shorter-time observation of a particular classical configuration, while the symmetric exact density represents an infinite-time average over equivalent configurations.

Continuing improvements in the accuracy and reliability of DFAs have produced an incremental reduction of self-interaction error. This work suggests that, for the reliable DFA description of strong correlation through symmetry breaking, full SIC must be achieved. Ref. \citep{Shahi2026Local} suggests that the most-important exact constraints for normal correlation are exactness for all one-electron and all uniform densities, but further work in that direction is needed. 

The harmonium two-electron example shows how the same physics can wear very different disguises: In a strongly-correlated state, symmetry preservation and symmetry breaking can be equivalent. Moreover, the ground-state energy can be computed and understood from at least three different perspectives: (a) As the expectation value of the Hamiltonian for a two-electron ground-state wavefunction that is analytically-expressible even when it is strongly-correlated. (b) As the sum of a semi-classical expansion in powers of $\hbar$. The zero-order term is the classical potential energy minimum, and the first-order term is the zero-point vibrational energy for the normal modes of electron vibration around classical equilibrium positions. (c) As the ground-state energy of non-interacting electrons in a Kohn-Sham mean field, corrected by the Hartree energy and by a self-interaction-free density functional for the exchange-correlation energy. Walter Kohn taught that the best way to understand something is to see it from at least three different perspectives.

\begin{acknowledgments}
    The work of CHW was supported by the U.S. Department of Energy, Office of Basic Sciences, under grant no. DE-SC0023356. JPP acknowledges support from that grant, and from the U.S. National Science Foundation under grant nos. DMR-2426275 and CHE-2533416.

    CL acknowledges funding support from the National Key Research and Development Program of China (Grant Nos. 2023YFA1507000 and 2025YFB3003603).

\end{acknowledgments}

\bibliography{apssamp1}

\end{document}